\documentclass[aps,prl,twocolumn,showpacs,amsmath,amssymb,groupedaddress]{revtex4-1}
\usepackage{graphicx}
\usepackage{times}
\begin{document}
\title{Traveling Dark Solitons in Superfluid Fermi Gases}
\author{Renyuan Liao}
\author{Joachim Brand}
\affiliation{New Zealand Institute for Advanced Study and Centre for Theoretical Chemistry and Physics, Massey University, Private Bag 102904 NSMC, Auckland 0745, New Zealand}
\date{\today}

\begin{abstract}
Families of dark solitons exist in superfluid Fermi gases. 
The energy-velocity dispersion and number of depleted particles completely determines the dynamics of dark solitons on a slowly-varying background density. For the unitary Fermi gas we determine these relations from general scaling arguments and conservation of local particle number. We find solitons to oscillate sinusoidally at the trap frequency reduced by a factor of $1/\sqrt{3}$.
Numerical integration of the time-dependent Bogoliubov-de Gennes equation determines spatial profiles and soliton dispersion relations across the BEC-BCS crossover and proves consistent with the scaling relations at unitarity.

\end{abstract}
\maketitle

Dark solitons are elementary nonlinear excitations that play a key role in understanding complex dynamics of superfluids \cite{Chang08,Ginsberg05}.
Superfluid Fermi gases have only recently become  accessible experimentally and their nonlinear wave dynamics are largely unexplored \cite{STR08,GAR07}. These systems offer the intriguing possibility to tune between the perturbatively accessible regimes of  Bose-Einstein condensation (BEC) of preformed pairs and Bardeen-Cooper-Schrieffer (BCS) superfluidity and a strongly correlated regime of unitarity-limited interactions. While the existence and properties of dark solitons in the BEC regime can be inferred from the solutions of Gross-Pitaevskii (GP) mean-field theory and experiments with atomic BECs, it is an outstanding question what happens outside this regime.
So far, only numerical solutions for stationary dark solitons within Bogoliubov-de Gennes (BdG) mean-field theory have been available \cite{STR07}.

In this work, we report theoretical results supporting the existence and detailing the properties of a family of traveling (grey) solitons that are parameterized by their velocity of propagation $v_s$. We are aware of parallel efforts to understand soliton dynamics in trapped Fermi gases \cite{Trento} and to determine grey soliton profiles \cite{Colorado}. For the unitary gas, we find a closed analytic form of the energy-velocity dispersion relation that is fully determined from a set of general assumptions: 
(a) Upon adiabatic change of the environment, the soliton can adjust its dynamical state and consistently conserve locally both energy and particle number. 
(b) Energy and particle number vanish as $v_s$ approaches the speed of sound. 
(c) The superfluid order parameter has a well defined phase step across the soliton that also vanishes under the conditions of (b).

\begin{figure}[t]
{\scalebox{0.30}{\includegraphics[clip,angle=0]{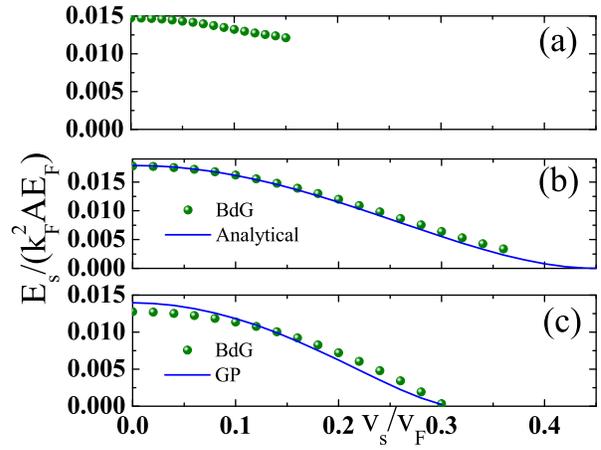}}}
\caption{(Color online) (a) Soliton energy $E_s$ as a function of velocity $v_s$ for different couplings: (a) $\eta=-0.5$ (BCS), (b) $\eta=0$ (unitary), and (c) $\eta=1$ (BEC), where the solid lines show the analytical relations (\ref{eq:eps}) and $E_s^{GP}=\frac{4}{3}\hbar n_Bc\left[1- v_s^2/c^2\right]^{3/2}$, respectively.}
\label{Fig5}
\end{figure}

\begin{figure}[t]
{\scalebox{0.30}{\includegraphics[clip,angle=0]{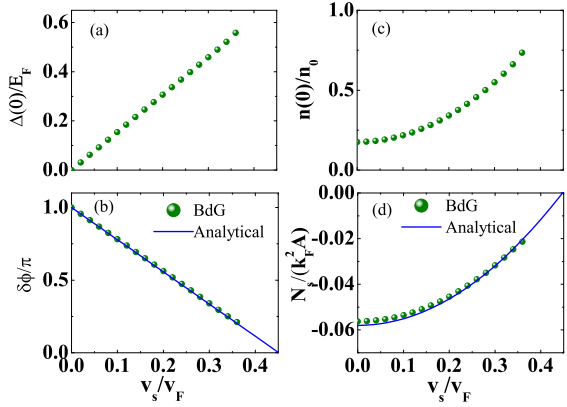}}}
\caption{(Color online) Soliton properties at unitarity $\eta=0$ as a function of velocity:
(a) the magnitude of the order parameter in the center, (b) the phase difference $\delta\phi = \arg[\Delta(\infty)]-\arg[\Delta(-\infty)]$, (c) the density in the center, and (d) the number of particles in the soliton. Dots show numerical data and lines plot Eqs.~(\ref{eq:phase}) and (\ref{eq:Ns}), respectively.}
\label{Fig4}
\end{figure}

Assumption (a) means that solitons can move adiabatically between regions of different  background density without disintegration and radiation. This non-trivial property is known to be true for GP solitons \cite{PIT04}. For the unitary gas the assumptions are supported by the excellent agreement found between the analytic dispersion and our numerical results based on BdG mean-field theory (see Figs.~\ref{Fig5}b and \ref{Fig4}b,d). The mean-field calculations further allow us to obtain spatial soliton profiles and dispersion relations for arbitrary interactions outside the unitarity limit. 

As a dynamical consequence we are able to predict oscillations of dark solitons in a harmonically trapped Fermi gas. While for BECs, the oscillation frequency was predicted \cite{ANG99} and observed \cite{becker2008oscillations,Weller08} to be reduced from the trapping frequency $\omega_t$ by a factor of $1/\sqrt{2}\approx 0.707$, we find the oscillation frequency further reduced across the BEC-BCS crossover. The BdG calculation yields $\omega/\omega_t$=0.480, 0.572, and 0.687 for $\eta=$ -0.5 (BCS regime), 0 (unitary) and 1 (BEC regime), respectively, where $\eta=1/(k_F a)$,  the Fermi wave number $k_F= (3 \pi^2 n)^{1/3}$ parametrizes the density $n$, and $a$ is the $s$-wave scattering length. Our analytic theory for the unitary case of  $\eta=0$ predicts $\omega/\omega_t = 1/\sqrt{3} \approx 0.577$, which is in excellent agreement with the numerical data.

Let us consider a superfluid Fermi gas with a soliton that is localized along the $z$ direction in a region small compared to the system length $L$ 
on a homogeneous background. 
We can extract the scaling of the system energy with density using the inverse Fermi wave number $k_F^{-1}$ and the Fermi energy $E_F=\hbar^2 k_F^2/(2m)$ as units of length and energy, respectively. 
We may express the grand canonical energy of the superfluid Fermi system $E' = \langle \hat{H}-\mu \hat{N}\rangle = [\varepsilon_h k_F L+ {\cal E} + {\mathcal {O}}((k_F L)^{-1})] k_F^2 A E_F$, where $A$ is the transverse area, $\varepsilon_h$ the dimensionless background energy and 
$E_s \equiv {\cal E} k_F^2 A E_F$
 is identified as the soliton energy. 
Physically, the soliton energy $E_s(\mu, v_s, a)$ depends on three independent parameters, where $v_s$ is the propagation velocity. Dimensionless ${\cal E}$, however, may only depend on the two dimensionless parameters of velocity $\tilde{v}=v_s/v_F$ and coupling strength $\eta$.  The dependence on the chemical potential $\mu$ is implicit through the density, which determines $k_F$. 

A special case arises for unitarity, where $a \to \infty$ and $\eta=0$ becomes independent of $\mu$. Anticipating that $E_s$ is an even function of the velocity, we consider ${\cal E}(\tilde{v}^2)$ as a function of $\tilde{v}^2$. Employing the equation of state of the unitary Fermi gas $\mu= (1+\beta)E_F$, where $\beta$ is the many-body parameter, we can extract the dependence on $\mu$
\begin{align}\label{eq:Es2}
E_s(\mu, v_s) = \mu^2 B {\cal E}(\tilde{v}^2),
\end{align}
where $\tilde{v}^2 = v_s^2 (1+\beta)m/(2\mu)$ and $B= 2Am/[(1+\beta)\hbar]^2$.
As a localized wave form, a soliton is characterized not only by its energy but also by its particle number $N_s = \int (n_s - n_0) d^3r = -\partial E_s/\partial \mu$, where $n_s$ is the soliton density and $n_0$ is the background density \cite{PIT04}. We find 
\begin{align}\label{eq:Ns}
N_s = v_s^2 \frac{m}{2}(1+\beta) B {\cal E}'(\tilde{v}^2) - 2 \mu B {\cal E}(\tilde{v}^2),
\end{align}
where ${\cal E}' =d{\cal E}/d\tilde{v}^2$. The particle number  and energy are thus completely determined by the same function ${\cal E}(\tilde{v}^2)$. We are now going to determine this function from the assumptions (a) -- (c).

Assumption (a) requires both $E_s$ and $N_s$ to be constants of the motion through inhomogeneous density. Since we are considering purely one-dimensional motion of the soliton, treating it as a quasiparticle, there is at most only a single independent constant of the motion. The condition that  $N_s$ and $E_s$ have identical contours in phase space leads to the condition 
\begin{align} \label{eq:Ncons}
\frac{\partial N_s}{\partial \mu}\frac{\partial E_s}{\partial v_s} = 
\frac{\partial N_s}{\partial v_s}\frac{\partial E_s}{\partial \mu} ,
\end{align}
where $\mu$ and $v_s$ represent the quasiparticle's coordinate and momentum, respectively. It follows from Eqs.~(\ref{eq:Es2}) and (\ref{eq:Ns}) that the third derivative of ${\cal E}$ vanishes identically and that ${\cal E}(\tilde{v}^2)$ can be parameterized by
\begin{align} \label{eq:eps}
{\cal E}(\tilde{v}^2) = \mathring{e}(\mathring{v}^2 -  \tilde{v}^2)^2 ,
\end{align}
where $\mathring{e}$ and $\mathring{v}$ are yet undetermined parameters.
The functional form (\ref{eq:eps}) already has important implications for soliton oscillations in a trapped gas: 

Requiring $d E_s/dt =0$, we find the Newtonian equation of motion for the soliton 
\begin{align} \label{eq:eom}
\left.\frac{d\mu}{dz} \right|_{z=z_s}- \frac{m (1+\beta)}{\mathring{v}^2}\ddot{z}_s = 0 .
\end{align}
In the case of harmonic trapping and under validity of the Thomas Fermi approximation, we can write $\mu(z) = \mu_0 - m \omega_t^2 z^2/2$ and Eq.\ (\ref{eq:eom}) reduces to a harmonic oscillator. The frequency $\omega/\omega_t = \mathring{v}/\sqrt{1+\beta}$ is independent of amplitude!

The parameter $\mathring{v}$ can be determined from assumption (b): From Eq.\ (\ref{eq:eps}) we find that the energy and particle number vanish when the dimensionless velocity $\tilde{v}$ reaches the critical value $\mathring{v}$. We expect this to happen at the speed of sound, which takes the value $c=\sqrt{(1+\beta)/3} v_F$. This leads to $\mathring{v}=\sqrt{(1+\beta)/3}$ and yields the oscillation frequency $\omega/\omega_t = 1/\sqrt{3}$. We have thus derived the oscillation frequency of a dark soliton in a harmonically trapped unitary gas from the assumptions (a) and (b).

The remaining coefficient  $\mathring{e}$ can be determined from the relation between the physical momentum of the soliton $p_s=mN_s v_s$ and the canonical momentum $p_c$, which is defined by $\partial E_s/\partial p_c|_\mu = v_s$. The difference between the two quantities accounts for the counterflow that would have to occur in a toroidal system to compensate for the phase difference $\delta\phi$ in the superfluid order parameter across the soliton \cite{shevchenko1988quasi}. For the superfluid Fermi gas the counterflow term was recently found by Pitaevskii \cite{Trento} 
\begin{align} \label{eq:Levs}
p_s-p_c = \hbar n_1 (\pi-\delta\phi)/2 ,
\end{align}
where $n_1 = n A = k_F^3 A/(3 \pi^2)$ is the one-dimensional density.
From Eqs.\ (\ref{eq:Es2}) and (\ref{eq:eps}) we evaluate the difference using $p_c= \int v_s^{-1} \partial E_s/\partial v_s\; {\rm d}v_s = -\hbar n_1 6 \pi^2 \mathring{e}\tilde{v}[\mathring{v}^{2}-\tilde{v}^2/3]$ to yield
\begin{align} \label{eq:our}
p_s-p_c = \hbar n_1 {4\pi^2 \mathring{e}}{\mathring{v}^2} \tilde{v} .
\end{align}
Comparing Eqs.\ (\ref{eq:our}) and (\ref{eq:Levs}), we find that the phase difference varies linearly with velocity in contrast to the GP soliton, where $\cos(\delta\phi^{\rm GP}/2) = v_s/c^{\rm GP}$. Fixing the remaining constant $\mathring{e}$ by requiring the phase step to vanish at the speed of sound [assumption (c)], we find
\begin{align}
\mathring{e} = \frac{\mathring{v}^{-3}}{8\pi}, \quad \delta\phi = \pi (1- v_s/c).
\label{eq:phase}
\end{align}
Thus, the energy and particle number dispersion (shown as full lines in Figs.~1b and 2d, repsectively) as well as the phase step for the family of dark solitons in the unitary gas (shown in Fig.~2b) are obtained without any free parameters. 
The success of this derivation shows, that the assumption (a) of particle-number conservation under quasiparticle motion is consistent with the universal scaling relations of the unitary Fermi gas. 

We have not yet proven that dark solitons exist. Within the realm of mean-field theory, this can be done by finding self-consistent solutions of the BdG equations. In addition to testing the stated assumptions against a physical theory, this allows us to determine spatial profiles as well as dispersion relations outside the unitary regime. 

We now more generally consider a Fermi gas with equal density for two spin components at zero temperature. The time-dependent BdG equations provide a convenient mean-field theory of the BEC-BCS crossover \cite{GAR07}
\begin{eqnarray}
  i\hbar\partial_t \begin{pmatrix}u_\nu(\mathbf{r},t)\\v_\nu(\mathbf{r},t)\end{pmatrix}=\begin{pmatrix}\hat{h} & \Delta(\mathbf{r},t)\\\Delta^*(\mathbf{r},t) & -\hat{h}\end{pmatrix}
   \begin{pmatrix}
    u_\nu(\mathbf{r},t)\\
    v_\nu(\mathbf{r},t)
   \end{pmatrix}\label{eq:eq3} ,
\end{eqnarray}
where $\hat{h}=\frac{\hbar^2}{2m}\nabla^2-\mu$ and $u$ and $v$ are space- and time-dependent quasi-particle amplitudes satsifying $\int d^3\bold{r}\left[u_\nu^*(\mathbf{r},t)u_{\nu'}(\mathbf{r},t)+v_{\nu}^*(\mathbf{r},t)v_{\nu'}(\mathbf{r},t)\right]=\delta_{\nu\nu'}$. The problem simplifies to a time-independent eigenvalue problem when we seek soliton solutions of the superfluid order parameter of the form 
$\Delta(z,t)=\Delta(z-v_s t)=\Delta(\xi)$
and write
$v_\nu(\mathbf{r},t) = (LA)^{-1/2}e^{i(p_x x+p_y y)-iE_{\mathbf{p},n}t/\hbar}v_{\mathbf{p},n}(\xi)$ and likewise for $u$. The energies $E_{\mathbf{p},n}$ are the eigenvalues of the resulting time-independent BdG equation, which contain the soliton velocity $v_s$ as a parameter. The transverse momentum $\mathbf{p}$ is discretized according to the transverse area $A$ of the computational box.
The above equations must be solved together with the equation for the order parameter
$\Delta(\xi)=-g\sum_{\mathbf{p},n}u_{\mathbf{p},n}(\xi)v_{\mathbf{p},n}^*(\xi)$ in a self-consistent way. The density is then given by 
$n(\xi)=2\sum_{\mathbf{p},n}|v_{\mathbf{p},n}(\xi)|^2$.
All sums are restricted to $0\le E_{\bold{p},n}\le E_c$ where $E_c$ is a high energy cutoff.
The coupling strength $g$ relates to the scattering length $a$ through the cutoff-dependent renormalization
$1/g=m/(4\pi\hbar^2a)-1/\Omega\sum_{\nu} 1/2\epsilon_{\nu}$~\cite{MRE93,YIP07} where $\epsilon_\nu$ is the energy of the state $\nu$ in the normal phase.
Open boundary conditions
are imposed by ensuring the hermiticity of the matrix and the proper symmetry of $\Delta(\xi)$ in a self-consistent way. We have implemented a generalized secant (Broyden's) method to find self-consistent solutions (with very small self-consistency error of $10^{-8}$ or better) from an initial profile with nontrivial phase structure.

\begin{figure}[t]
{\scalebox{0.30}{\includegraphics[clip,angle=0]{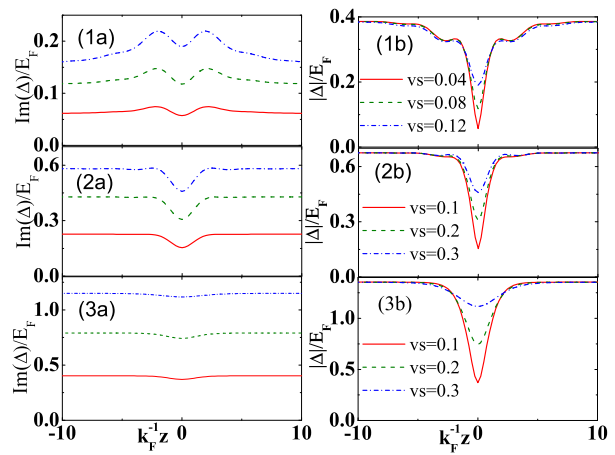}}}
\caption{(Color online) The spatial structure of the soliton order parameter $\Delta$ at different velocities by its imaginary part (a) and magnitude (b) in the BCS regime at $\eta=-0.5$ (1a,b), unitarity limit $\eta=0$ (2a,b), and BEC regime $\eta=1$ (3a,b).}
\label{Fig2}
\end{figure}

\begin{figure}[t]
{\scalebox{0.30}{\includegraphics[clip,angle=0]{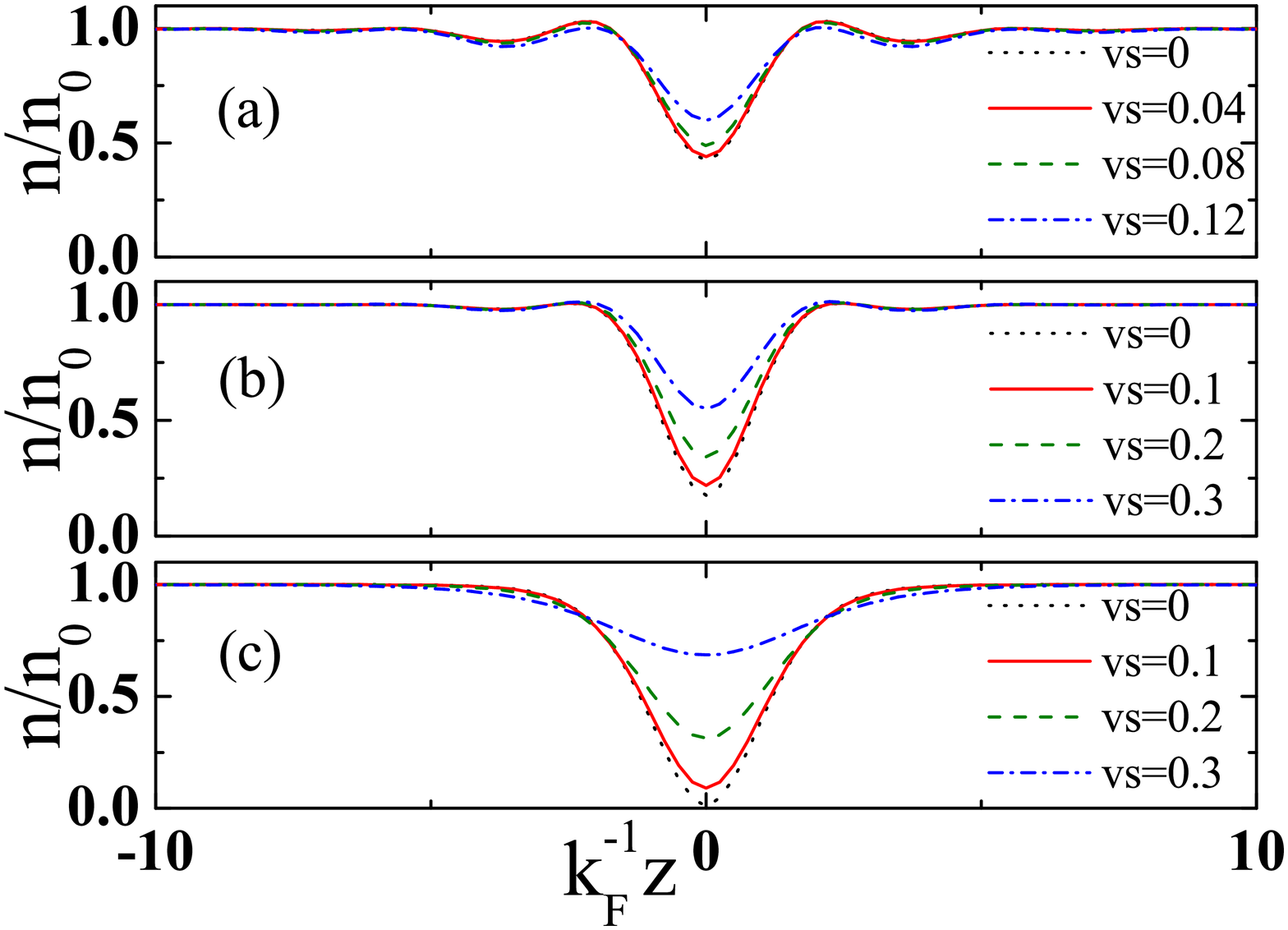}}}
\caption{(Color online) The density profile at different velocities for (a) $\eta=-0.5$ (BCS), (b) $\eta=0$ (unitarity) and (c) $\eta=1$ (BEC).}
\label{Fig3}
\end{figure}

The spatial structure of the dark soliton solutions is shown in 
Figs.\ \ref{Fig2} and \ref{Fig3} for three values of the interaction parameter $\eta$. The main feature is a density notch and dip in the order parameter, which become shallower with increasing velocity for all interacting regimes. In the BEC regime, the profiles most closely resemble the GP dark solitons with constant imaginary part (with appropriate choice of a global phase) and bell-shaped, $\text{const} - \text{sech}^2(z/\ell)$, density notch. This is expected as the BdG equation converges toward the GP equation for large positive $\eta$ \cite{STR03}. The imaginary part of the order parameter clearly develops structure for finite velocities in the unitarity and BCS regime, which is a striking new feature of the BdG grey solitons as seen in panels (1a) and (2a) in Fig.~\ref{Fig2}.

While the length scale for the BEC soliton is $\ell=\hbar/(m\sqrt{c^2-v_s^2})$ from GP theory, there is no clear evidence in our data for a velocity dependence of the length scale in the unitarity limit. There we expect on general grounds that the only length scale is $k_F^{-1}$. In the BCS regime, we expect small scale Friedel oscillations with size $2k_F^{-1}$, and a second length scale in the Cooper pair size $\xi_C=\hbar v_F/\Delta_0$, which evaluates to about $5 k_F^{-1}$ for $\eta=-0.5$ \cite{STR07}. From Fig.~\ref{Fig2}, panel (1a) and our experience with boundary effects, it appears that there is an additional velocity dependence and the total size grows with increasing velocity.

Relevant velocity scales for the problem are the speed of sound $c = \sqrt{n (\partial\mu/\partial n)/m}$ and the pair breaking velocity $mv_{sp}^2=\sqrt{\mu^2+\Delta_0^2}-\mu$. We consistently found it difficult to converge to self-consistent solutions approaching these velocities from below and thus assume that soliton solutions exist only below $v_c=\min(c,v_{sp})$, which is also the critical velocity for dissipationless motion of infinitesimal impurities \cite{STR08,Spuntarelli2010111}. $v_c$ takes a maximum around unitarity, where $c\approx v_{sp}$. For $\eta \gtrsim 0$ (BEC to unitarity), $c<v_{sp}$ and thus solitons are limited by the speed of sound. In the BCS regime, pair breaking dominates and $v_{sp}$ limits soliton propagation.

The dimensionless soliton energy as a function of the velocity is shown in Fig.~\ref{Fig5}. In all three regimes the energy is positive with negative curvature, which supports the understanding of a dark soliton as a quasiparticle with negative effective mass. In the unitarity limit, the numerical data fits beautifully with the analytical result (\ref{eq:eps}), where we have used $1+\beta=0.6082$ consistent with our BdG calculation of the chemical potential. The GP formula for the energy shown in panel (c) is strictly valid only in the limit of large $\eta$, which explains the small deviations from the numerical data.

Additional properties of dark solitons are shown in Fig.~\ref{Fig4} for the unitarity limit. The order parameter at the soliton center shown in panel (a) intriguingly appears to vary linearly with velocity, as would be the case for GP solitons.
The phase step and particle number agree very well with the analytical predictions based on Eqs.\ (\ref{eq:eps}) and (\ref{eq:Ns}). We have further tested the assumption (a) that $N_s$ is a constant of the motion by directly checking Eq.~(\ref{eq:Ncons}), which is fulfilled within our expected numerical errors.

Finally, we consider small amplitude soliton oscillations in a harmonically trapped Fermi gas beyond the unitarity regime. From the Thomas-Fermi and local density approximation we have $E_s(v_s,\mu_0-m\omega_t^2 z_s^2/2) = \text{const}$ and taking a time derivative obtain 
 \begin{eqnarray}
       \frac{1}{v_s}\frac{\partial E_s}{\partial v_s}\ddot{z}_s +N_s m \omega_t^2 z_s = 0.  \label{eq:sol}
 \end{eqnarray}
 Noting that both $E_s$ and $N_s$ should be even functions of velocity, Eq.~(\ref{eq:sol}) describes a harmonic oscillator.
Defining the effective mass $M_s = v_s^{-1}\frac{\partial E_s}{\partial v_s}|_{v_s=0}$, we find the frequency of small oscillations 
  \begin{eqnarray}
   \frac{\omega^2}{\omega_t^2}=\sqrt{\frac{mN_s(0)}{M_s}} ,
 \end{eqnarray}
which is nicely interpreted as the ratio between the physical mass $m N_s$ and effective mass $M_s$.
Both $N_s(0)$ and $M_s$ are negative and can be easily extracted from our numerical data. We find the oscillation frequencies $\omega/\omega_t$=0.480, 0.572, and 0.687 for $\eta$=-0.5, 0 and 1 respectively.

The oscillation frequency of solitons thus decreases significantly from the BEC towards the BCS regime in agreement with time-dependent simulations~\cite{Trento}. In the unitarity regime we were able to determine the dispersion relation in closed form starting from a small number of global assumptions. In particular we find that the oscillation frequency does not depend on
the many-body parameter $\beta$ in the equation of state or other details of the soliton solutions. Dark solitons thus offer the opportunity to explore complimentary properties of the unitary gas to previous studies that pinpointed the equation of state \cite{STR08}. The important questions of stability of dark solitons against strong quantum fluctuations and the 
consistency of local energy and particle number conservation 
deserve to be studied beyond mean-field theory and thus make an excellent subject for future experimental investigation.

We acknowledge helpful discussions with L.\ Carr, F.\ Dalfovo, T.\ Ernst, L.\ Pitaevskii, R.\ Scott, and S.\ Stringari. J.B.\ received support from the Marsden Fund of New Zealand (Contract No. MAU0910).

\end{document}